\documentclass[aps,prl,twocolumn,showpacs,10pt,floatfix,superscriptaddress]{revtex4-1}
\usepackage[dvips]{graphics}
\usepackage{graphicx}
\usepackage[utf8]{inputenc}
\usepackage{amssymb}
\usepackage{epstopdf}
\usepackage{bm}
\usepackage{dcolumn}
\usepackage{epsfig}
\usepackage{latexsym}
\usepackage{amsmath,mathtools}
\usepackage{color}
\usepackage{array}
\usepackage{enumerate,dsfont}

\def\XXint#1#2#3{{\setbox0=\hbox{$#1{#2#3}{\int}$ }
\vcenter{\hbox{$#2#3$ }}\kern-.5\wd0}}

\newcommand{\ket}[1]{\left |\mbox{$#1$}\right\rangle}

\makeatletter
\newcommand*{\rom}[1]{\expandafter\@slowromancap\romannumeral #1@}
\makeatother

\begin{document}

\title{From Coupled Rashba Electron and Hole Gas Layers to 3D Topological Insulators}

\author{Luka Trifunovic}
\affiliation{Department of Physics, University of Basel, Klingelbergstrasse 82,
CH-4056 Basel, Switzerland}
\affiliation{Dahlem Center for Complex Quantum Systems and Physics Department, Freie Universit\"at Berlin, Arnimallee 14, 14195 Berlin, Germany}
\author{Daniel Loss}
\affiliation{Department of Physics, University of Basel, Klingelbergstrasse 82,
CH-4056 Basel, Switzerland}
 \author{Jelena Klinovaja}
\affiliation{Department of Physics, University of Basel, Klingelbergstrasse 82,
CH-4056 Basel, Switzerland}
\date{\today}

\begin{abstract}
We introduce a system of  stacked two-dimensional electron and hole gas layers with
Rashba spin orbit interaction and show that  the tunnel coupling between the layers
induces a strong three-dimensional (3D) topological insulator phase.
 At each
of the two-dimensional bulk boundaries we find the spectrum consisting of a single anistropic Dirac cone,
which we show by analytical and numerical calculations. Our setup has
a unit-cell consisting of four tunnel coupled Rashba layers and presents a
 synthetic strong 3D topological insulator and is distinguished by its
rather high experimental feasibility.
\end{abstract}

\maketitle
{\it Introduction.} Since the discovery of the quantum Hall effect there has been
immense theoretical interest focused on understanding topological phases of
quantum matter~\cite{QHE,HasanRMP}.
The interest was not solely concentrated on classification of these novel
phases~\cite{classification_t0p}, which goes beyond the Landau paradigm of phase
transitions, but also on potential applications of the topologically ordered
phases, in particular for storing quantum information in a manner that is resilient
to local imperfections~\cite{Kitaev20032}. Additionally, the electronic surface 
states of a strong topological insulator (TI)~\cite{pankratov,HasanRMP},
being an example of a 3D topological phase of matter, forms
a  two-dimensional (2D) topological metal, which is 'half' of an ordinary
metal~\cite{HasanRMP}. Such 2D topological metals are notable for the fact that
their electrons cannot be localized even in presence of strong disorder, as long
as the bulk energy gap of the parent strong 3D TI is
intact~\cite{PhysRevLett.99.146806}.

There are strong indications that  certain materials, such as semiconducting alloys,
behave as strong 3D TIs~\cite{HasanRMP}. Despite great
success in this field, both theoretically and experimentally, there are still
certain issues that need to be resolved, in particular that
strong TIs suffer from bulk conduction due to
chemical imperfections. 
 Thus, there is a strong need for synthetic materials
where one has enough control over the system parameters in order to achieve a
topological phase with a sufficiently large bulk gap which excludes bulk conduction.

One of the very successful approaches for theoretically constructing 2D
topological phases of matter is using anisotropic hopping or a coupled wire
construction~\cite{PhysRevLett.58.270,PhysRevB.51.3285,PhysRevLett.88.036401,PhysRevLett111.196401,PhysRevB.89.085101,epb,tobias_40,coupledWKY,coupledWNeupert,coupledWOreg,PhysRevB.91.085426,santos_50,sagi_41,stoudenmire_43,teo}.
Apart from being very intuitive, this approach allows non-perturbative
treatment of the electron-electron interactions and is thus suitable for study
of fractional topological phases.  Recently, a strong effort was made to extend
this approach to the study of 3D TIs, where topological phases related to  weak
TIs were obtained, as well as Weyl semimetal
phases~\cite{2015arXiv150601364M,oreg,teo}. Despite the great theoretical
insight this approach gives, its main drawback in the case of 3D systems is
that the resulting setups are rather complex and thus not easy to realize
experimentally. 
\begin{figure}
\includegraphics[width=\columnwidth]{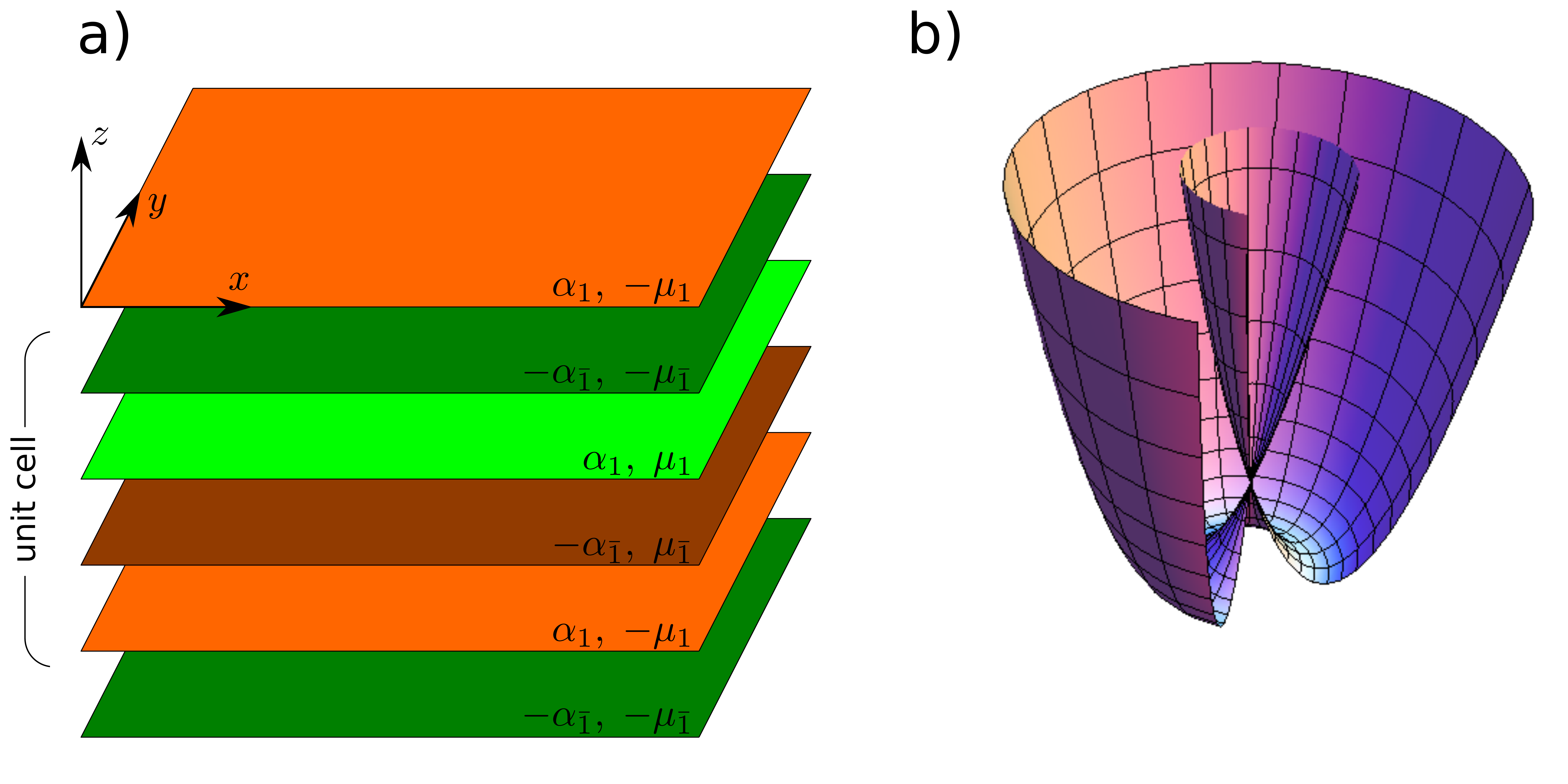}
\caption{
Panel a) shows the setup consisting of a stack of  layers arranged in the
$xy$-plane and tunnel coupled along the $z$ axis. The layers colored in green
(orange) denote electron (hole) 2DEGs with Rashba SOI and at chemical potential
$\mu_{\tau}$. The brightness of the color encodes two possible values of the
Rashba SOI $ \alpha_{\tau}$. Panel b) shows the dispersion of a 2DEG with Rashba SOI.}
\label{fig:layers}
\end{figure}
In this paper we take a different approach, instead of coupled
wires~\cite{PhysRevLett.58.270,PhysRevB.51.3285,PhysRevLett.88.036401,PhysRevLett111.196401,PhysRevB.89.085101,epb,tobias_40,coupledWKY,coupledWNeupert,coupledWOreg,PhysRevB.91.085426,santos_50,sagi_41,stoudenmire_43,teo,2015arXiv150601364M}
we introduce a construction of coupled 2D layers, see Fig.~\ref{fig:layers}.
Each layer is a simple 2D electron gas (2DEG) with Rashba spin-orbit
interaction (SOI)~\cite{rashba}. By generalizing the coupled wires
approach~\cite{PhysRevB.91.085426} to coupled layers we arrive at a rather
simple realization of a strong 3D TI. 

{\it Model.} We consider a system consisting of tunnel coupled layers of 2DEGs
stacked along the $z$-axis, see Fig.~\ref{fig:layers}. In each 2DEG  we include
SOI and we assume it to be of Rashba type~\footnote{Our results still hold if
Rashba  is replaced by Dresselhaus SOI. In case when both Rashba and
Dresselhaus SOI are present, our scheme still works if one of them dominates.}.
In our model, we work with two different values of SOI that could be chosen
almost arbitrary (see below) and do not require special tuning.  In contrast to
that, the chemical potential $\mu_{\tau}$ in each layer should be individually
tuned  to the value determined by the corresponding SOI. Our setup has a unit
cell consisting of four Rashba 2DEG layers. 

\begin{figure}
\includegraphics[width=\columnwidth]{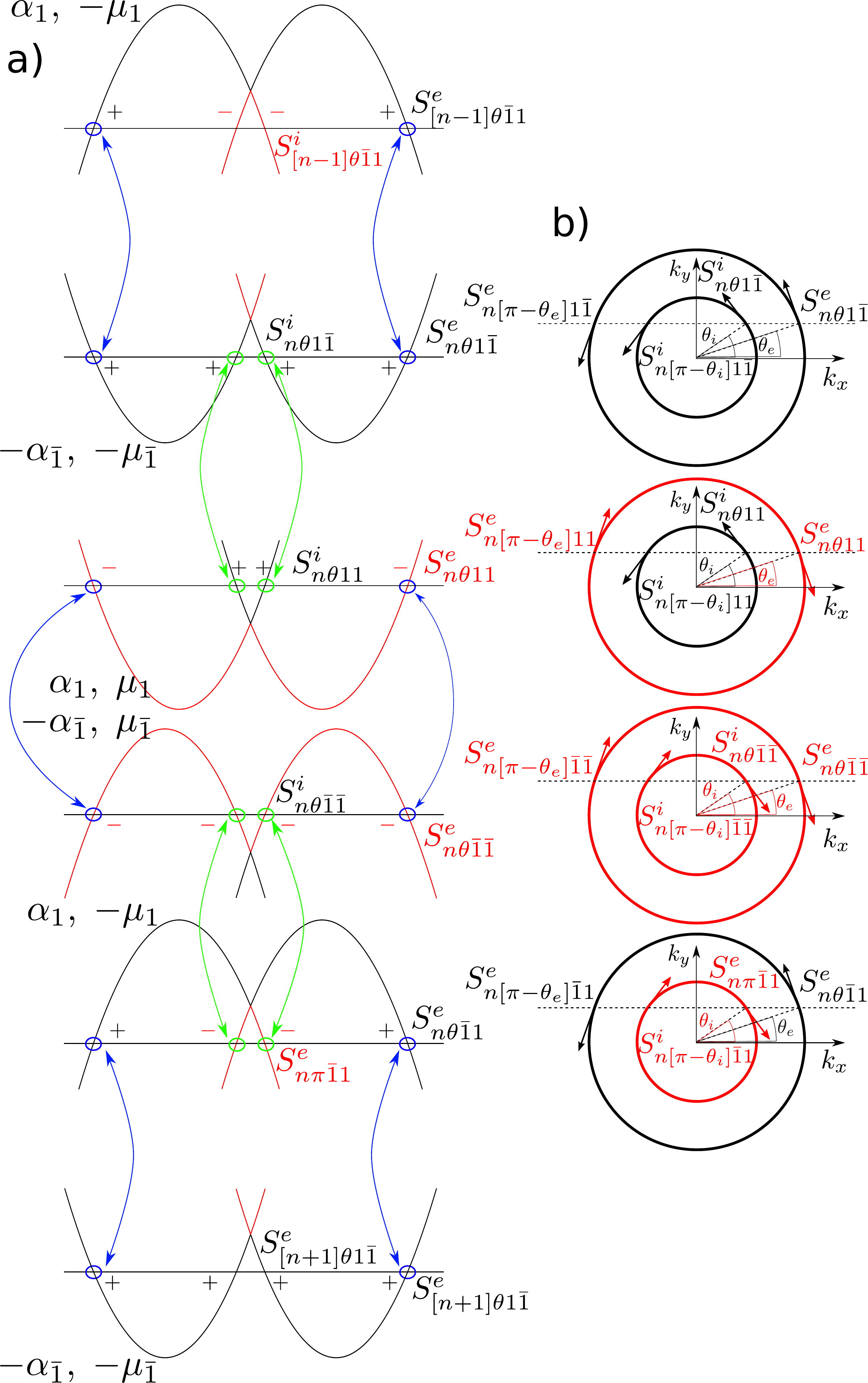}
\caption{Panel a) shows the dispersion relation of each layer for fixed
  $\theta$. The chemical potentials $\mu_\tau$ are chosen such that inner and
  outer Fermi surfaces have the same radii across  different layers. The arrows
  indicate where the tunneling between the layers opens up gaps (small green
  circles). Note that the bottom and top layers stay gapless and have a
  dispersion consisting of a single Dirac cone with spin locked to momentum due
to time reversal invariance. Panel b) shows the interior and exterior Fermi surface of each layer
with the cuts for $k_y=const$. The fields for interior (exterior) left and
right movers $S_{n\theta_i\eta\tau}^i,S_{n[\pi-\theta_i]\eta\tau}^i$
($S_{n\theta_e\eta\tau}^e,S_{n[\pi-\theta_e]\eta\tau}^e$) have in general
  different spin orientations.}
\label{fig:TI}
\end{figure}

A single 2DEG layer with Rashba SOI is described by the following
Hamiltonian~\cite{rashba} 
\begin{align}
H_0&=- \hbar^2{(\partial_x^2+\partial_y^2)}/2m_0-i \alpha (\sigma^x \partial_y - \sigma^y \partial_x),
\label{eq:2DEGSOI}
\end{align}
where $\alpha$ is the strength of the Rashba SOI and $m_0$  the electron mass
in the given band. We can diagonalize the above Hamiltonian by taking the local
spin quantization axis $\bm s = (-\sin \theta, \cos \theta)$ to be always
perpendicular to the momentum $\bm k = (k_x, k_y)\equiv k (\cos \theta, \sin
\theta)$,
\begin{align}
E_{\mp}(\bm k)&=\hbar^2 k^2/2m_0\mp\alpha k,
\label{eq:2DEGSOIp}
\end{align}
where the  upper (lower) sign corresponds to the spin orientation being along
(opposite to) $\bm s$ chosen for $\alpha>0$ and to the lower (higher) energy for
a fixed $\bm k$, where the corresponding spinors are given by
\begin{align}
 {\ket {\mp; \theta}} &=\frac{1}{\sqrt{2}}
  \begin{pmatrix}
    1\\
    \pm i e^{i  \theta}
  \end{pmatrix}.
\label{eq:Sdef}
\end{align}
We note here that the spin orientation is clockwise (anticlockwise) for ${\ket
{+; \theta}}$ (${\ket {-; \theta}}$).  The dispersion relation
Eq.~(\ref{eq:2DEGSOIp}) is depicted in Fig.~\ref{fig:layers}b, and the shape of
the Fermi surfaces and the spin orientations in
Fig.~\ref{fig:TI}b. 

The setup we consider herein consists of four stacked layers
composing the unit cell,
which then periodically repeats in  $z$-direction with  spacing
$a_z$ between layers. Each of the four layers of the unit cell is
labeled by two indices $\eta=\pm 1$ and $\tau=\pm 1$. The index $\eta=1$
($\eta=\bar 1$) corresponds to an electron (hole) dispersion relation. The index
$\tau$ refers to two different values of the SOI, $\alpha_1$ and $\alpha_{\bar
1}$, where without loss of generality we assume that
$0<\alpha_1<\alpha_{\bar1}$. The ordering of the layers inside the unit cell is
shown in Fig. \ref{fig:TI}a. Two electron layers are followed by two hole layers
as the SOI magnitude $\alpha_\tau$ alternates from layer to  layer.

The total Hamiltonian of the system is $H=\sum_{n=1}^N\int dxdy\ {\cal
H}_n(x,y)$, where $N$ is the total number of unit cells and the Hamiltonian
density  is given by ${\cal H}_n={\cal H}_{n0}+{\cal H}_{nt}$ with ${\cal
H}_{n0}=\sum_{\{\tau,\eta =1,\bar 1\}} {\cal H}_{n\eta \tau}$, where
\begin{align}
  \label{eq:Hcell}
  {\cal H}_{n\eta \tau} = &\sum_{\sigma, \sigma'}
   \Psi^\dagger_{n\eta \tau\sigma}\Big[-\frac{\eta \hbar^2}{2m_0} (\partial_x^2+\partial_y^2) +\eta \tau \mu_{\tau} \nonumber\\
  &- i\tau \alpha_\tau (\sigma^x \partial_y - \sigma^y \partial_x)
  \Big]_{\sigma \sigma'}\Psi_{n\eta \tau\sigma'}.
\end{align}
The electron (hole) annihilation operator $\Psi_{n\eta \tau\sigma} (\bm r)$
acts on particles with spin $\sigma$  at the position $\bm r=(x,y)$ of the $(n
\eta \tau)$-layer. The chemical potential $\mu_\tau$ is calculated from the
crossing point at $k=0$ determined by the SOI energy $E_{so,\tau}=\hbar^2
k_{so,\tau}^2/2m_0$ with the SOI wavevector $k_{so,\tau} = m_0
\alpha_\tau/\hbar^2$. The dispersion relation (for fixed $\theta$) of each
layer is shown in Fig.~\ref{fig:TI}a and can be easily generalized to all
directions of $\bm k$.  In the following, we fix the chemical potentials as
$\mu_1=\mu_{\bar 1}=E_{so,\bar1}-E_{so,1}$. This choice ensures that the
interior (exterior) Fermi surfaces have the same radius
$k_{Fi}=k_{so,\bar1}-k_{so,1}$ ($k_{Fe}=k_{so,\bar1}+k_{so,1}$) across all the
layers. Additionally, we need to assume that $\mu_\tau\gg t$.

The tunneling between the layers is assumed to be spin-independent and takes
the following form,
\begin{align}
{\cal H}_{nt}&= \sum_{\sigma
\langle \tau \eta;n^\prime \tau^\prime  \eta^\prime\rangle
} t \Psi^\dagger_{n\eta \tau \sigma}(\bm r)\Psi_{n^\prime
 \eta^\prime \tau^\prime\sigma}(\bm r)+\mathrm{H.c.},
\label{eq:Hnt}
\end{align}
where the summation runs over all neighbouring layers.

First, we demonstrate that the top and bottom layers host gapless modes with
a helical Dirac spectrum. For the moment, we assume that the system is infinite
and  translationally invariant in $x$- and $y$- directions and we introduce
momenta $k_x$ and $k_y$, which are good quantum numbers. Alternatively, due to
rotation invariance, one can change to polar coordinates 
with momenta $k_r$ and $k_\theta$. This allows us to treat the
problem as effectively one-dimensional if the orbital degree of freedom is
integrated out, see Fig.~\ref{fig:TI}. The wavefunction can be represented
close to the Fermi surface in terms of slowly varying fields $S_{n \theta\eta
\tau }^{e/i}$,
\begin{align}
 \Psi_{n\theta  \eta \tau  \sigma}(x,y)=& \sum_{\delta=e,i}
\alpha_{\delta \theta\eta\tau\sigma} S_{n \theta\eta \tau}^\delta e^{i k_{F\delta} (x\cos\theta+y\sin\theta) }  
\label{eq:linearization_11}
\end{align}
with the angle $\theta \in [0, 2\pi)$, $\alpha_{e\theta\eta\tau\sigma}=\langle
\sigma \vert -\tau\cdot\eta,\theta\rangle$  and
$\alpha_{i\theta\eta\tau\sigma}=\langle \sigma\vert \eta,\theta\rangle$. The
kinetic term can be rewritten as
\begin{align}
\tilde H_{0} = & - i \hbar  \sum_{\mathclap{\substack{\delta=i,e\\ 
 n\theta;\tau,\eta = \pm 1}}} \beta_\delta\eta
 \upsilon_{F\tau}   (S_{n \theta\eta \tau }^\delta)^\dagger
 \frac{\partial}{\partial r} S_{n \theta\eta \tau }^\delta,
\end{align}
where $\beta_e=1$ and $\beta_i=\tau$. We also take into account that the Fermi
velocities $\hbar \upsilon_{F\tau} = \partial E_\tau/\partial k |_{\mu_\tau} $
are different. The tunneling terms induce couplings between interior/exterior
Fermi surfaces of different layers,
\begin{align}
\tilde H_t = t& \sum_{n\theta} [(S_{n \theta 11 }^e)^\dagger  
S_{n \theta \bar 1 \bar1 }^e +
(S_{[n-1] \theta \bar 11 }^e)^\dagger S_{n \theta  1 \bar1 }^e\nonumber\\
& + (S_{n \theta 1 \bar 1 }^i)^\dagger S_{n \theta 1 1 }^i + (S_{n \theta \bar 1 \bar 1 }^i)^\dagger S_{n \theta \bar 1 1 }^i ]+ H.c. \label{tun1}
\end{align} 
Here, we keep only non-oscillating terms and take into account the spin
conservation during the tunneling, see Fig.~\ref{fig:TI}a. Importantly, all
coupling terms in Eq. (\ref{tun1}) involve fields with opposite signs  of Fermi
velocities and each field, exept for the ones belonging to the top  and
bottom layers, has a partner to which it is coupled. This results in the
opening of gaps at the Fermi level such that the bulk spectrum is fully gapped.
However,  the exterior Fermi surface field $S^e_{1 \theta 1 \bar 1}$ of the 
top-most layer  and the interior Fermi surface field $S^i_{N \theta \bar 1
1}$ of the bottom-most layer do not have partners in Eq.~(\ref{tun1}) and,
thus, stay gapless as all the remaining layers are insulating. As was noted
above, $S^e_{1 \theta 1 \bar 1}$ and $S^i_{N \theta \bar 1 1}$  describe the
helical Dirac cones in which spin direction is locked to the momentum direction.
In our case, the spin direction stays always perpendicular to the momentum,
see Fig. \ref{fig:TI}b. Such surface states are the hallmark of a strong 3D
TI~\cite{HasanRMP}. 

Since the rotational symmetry is broken, it is far from obvious that the
surface states exist on \textit{any} 2D boundary. To this end, we demonstrate
that helical surface states also exist if a hard-wall boundary is added, say, at the
plane $x=0$. To this end we assume that the system is infinite in $y$- and
$z$-direction. Since the system is translation invariant in $y$-direction
($z$-direction), $k_y$ ($k_z$) is a good quantum number defined via
$\Psi_{k_z}=\Sigma_ne^{ink_za}\Psi_{ n}/\sqrt{N}$, where $a=4a_z$ is the
unit-cell size. The $y$-dependence of the total wavefunction is given trivially
as $\Psi_{k_yk_z\eta\tau\sigma}(x,y)=e^{ik_y y}\Psi_{k_yk_z\eta\tau\sigma}(x)$.
Since both $k_y$ and $k_z$ are good quantum numbers the problem is effectively
one-dimensional, see Fig.~\ref{fig:TI}b. To simplify the problem further, we
linearize the motion in the $x$-direction which is achieved with the ansatz
following from Eq.~(\ref{eq:linearization_11}),
\begin{align}
 \Psi_{k_yk_z\eta\tau\sigma}(x)&=\sum_{\mathclap{\substack{\delta=i,e\\ 
 \theta\in\{\theta_\delta,\pi-\theta_\delta\}}}}
\hspace{9pt} \alpha_{\delta\theta\eta\tau\sigma} S^\delta_{k_z\theta\eta\tau}(x)e^{ik_{F\delta}x\cos\theta},
  \label{eq:linearization}
\end{align}
where $S^\delta_{k_z\theta\eta\tau}$ is the Fourier transform of
$S^\delta_{n\theta\eta\tau}$. The above
ansatz~\cite{linearization2a,linearization2b,linearization2c} is valid for
$k_y<k_{Fi}$ and $t\ll\vert E_{so,\tau}-\hbar^2(k_y-\tau k_{so,\bar\tau})^2/2m\vert$ with
$E_y=\hbar^2k_y^2/2m_0$. The angles $\theta_i$ and $\theta_e$ are defined in
Fig.~\ref{fig:TI}b or explicitly expressed by
$\cos\theta_\delta=\sqrt{k_{F\delta}^2-k_y^2}/k_{F\delta}$. The spin
orientation is determined by $\alpha_{\delta\theta\eta\tau\sigma}$ and depends
on $\theta_\delta$ which in turn depends on $k_y$, see Fig.~\ref{fig:TI}b.

After performing  above
linearization~\cite{linearization2a,linearization2b,linearization2c}, we arrive
at the effective Hamiltonian
\begin{align}
   &\bar   H_0=-i 
   \sum_{\delta=i,e}\hbar\hspace{5pt}\sum_{\mathclap{\substack{\eta,\tau=\pm1\\ 
   \theta\in\{\theta_\delta,\pi-\theta_\delta\}}}} \hspace{6pt}\beta_\delta\eta
   \upsilon_{F\tau} \cos\theta  (S_{k_z\theta\eta\tau}^\delta)^\dagger
   \partial_x S_{k_z\theta\eta\tau}^\delta, \\
&\bar H_t = t \sum_{\mathclap{\theta\in\{\theta_i, \pi-\theta_i\}}}\left[ 
  (S_{k_z  \theta1 \bar 1 }^i)^\dagger S_{k_z  \theta_i 1 1 }^i + (S_{k_z  \theta\bar 1 \bar 1 }^i)^\dagger S_{k_z  \theta_i \bar 1 1 }^i\right]  \label{tun} \\
&+t\sum_{\mathclap{\theta\in\{\theta_e, \pi-\theta_e\}}}\left[ (S_{k_z \theta 11 }^e)^\dagger S_{k_z  \theta \bar 1 \bar1 }^e +
  e^{ik_z a}(S_{k_z \theta_e \bar 11 }^e)^\dagger S_{k_z \theta_e  1 \bar1 }^e\right]
+ H.c. \nonumber
\end{align}
It is readily noticeable from Fig.~\ref{fig:TI}a, that the Hamiltonian breaks
down into $2\times2$ blocks, formed by the fields coupled by the tunneling.
After inserting the ansatz $S^\delta_{k_z\theta\eta\tau}(x)\sim e^{ik_\delta
x}$, we arrive at the bulk spectrum around the interior and exterior Fermi
surfaces,
\begin{align}
  E_{\delta,\pm}=&k_{\delta}(\upsilon_1-\upsilon_{\bar1})\cos\theta_\delta\nonumber\\
  &\pm\sqrt{4t^2+k_{\delta}^2(\upsilon_1+\upsilon_{\bar1})^2\cos^2\theta_\delta},
   \label{eq:bulkspectrume}
\end{align}
where $k_{\delta}=k_x- k_{F\delta}\cos\theta_\delta$ and $\delta=e,i$. The bulk
spectral gap is given by
$\Delta=2t\sqrt{\upsilon_1\upsilon_{\bar1}}/(\upsilon_1+\upsilon_{\bar1})$.
The dispersion relation is determined by
\begin{align}
  \sin\left( 2\Omega
  \right)=\pm\frac{2\sin(k_za/2)\cos\theta_e\cos\theta_i}{\cos\theta_e+\cos\theta_i},
  \label{eq:dispersion}
\end{align}
and plotted in the SM. We note that
$E(k_y,k_z=0)$ is independent of $k_y$, which results in degeneracy. This
degeneracy is due to fact that we only retained resonant processes in our
perturbation analysis \cite{KSL2012}. If the problem is solved numerically (see below), this accidental degeneracy is lifted except at $\bm k=0$, where it is
protected by time reversal symmetry. Also any perturbation in the
chemical potentials lifts such a degeneracy and one is
left with an single anisotropic Dirac cone. To demonstrate this
explicitly, we assume a detuning $\delta\mu$ of chemical potential in the first
layer. For each value of $k_y$ there is a twofold degeneracy which is lifted by
such a perturbation. After performing the perturbation expansion within the
twofold degenerate subspace we arrive at the following dispersion relation
\begin{align}
  E(k_y,k_z=0)&=\frac{\delta\mu}{8}\left( 1-\frac{k_{Fi}}{k_{Fe}}
  \right)\frac{k_y}{k_{Fi}},
  \label{eq:dispersionDirac}
\end{align}
where we assumed $\upsilon_1=\upsilon_{\bar1}=\upsilon$, $k_y\ll k_{Fi}$, and
$t\ll k_{Fi}\upsilon$, see the SM~\cite{SM} for details.

\begin{figure}
  \includegraphics[width=0.8\columnwidth]{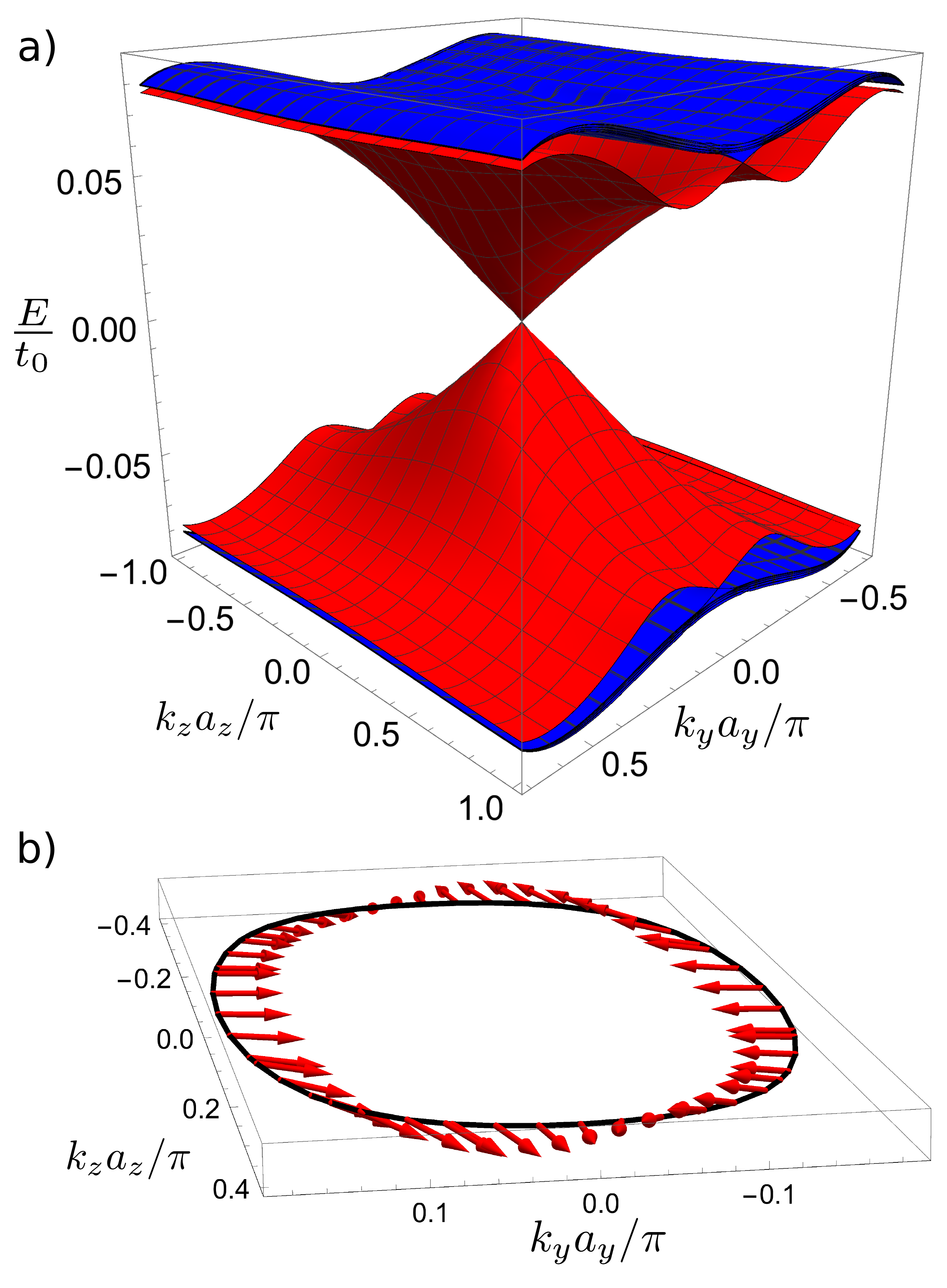}
  \caption{a) Dispersion relation of the surface states (red) localized in the
    $yz$-plane as well as bulk states (blue)  obtained numerically, see Eq.
    (\ref{tb}). At small momenta, the surface states form a single anisotropic Dirac
    cone, but merge with the bulk states at large momenta. b) The spin
    orientation (red arrows) of the first layer of the unit-cell for a fixed
    energy $E/t_0=-0.03$ and at the position $x_0=3a_x$ away from the
    left edge. The parameter values assumed are: $\bar \alpha_{1}=0.3 t_0$,
  $\bar \alpha_{\bar1}=0.55 t_0$, $\vert\mu_1\vert=\vert\mu_{\bar1}\vert=0.2125 t_0$, and
  $t=0.1 t_0$. The spin orientation is locked to the momentum direction confirming the
strong TI phase.}
  \label{fig:num}
\end{figure}

We finally address the above model numerically and study the edge states along
the $yz$ layer in the tight-binding model framework with $k_y$ and $k_z$ being
good quantum numbers. The corresponding tight-binding Hamiltonian is given by
$H=\sum_{k_y k_z \eta \tau} H_{0k_y k_z\eta \tau} + \sum_{k_y k_z} H_{tk_yk_z}$
with
\begin{align}
  &H_{0k_yk_z\eta \tau} = -
  \sum_{n \sigma} \big[  \eta ( t_0 \cos ( k_y a_y) +\mu_\tau/2) c_{k_yk_z\eta \tau
  n \sigma}^\dagger c_{k_y k_z\eta \tau n \sigma} \nonumber \\ 
  &- \eta t_0  c_{k_yk_z \eta \tau
  (n+1) \sigma}^\dagger c_{k_y k_z\eta \tau n \sigma} \big]\nonumber \\
&+  \bar \alpha_\tau \sum_{{n}} [c_{k_y k_z\eta \tau (n-1) \bar 1}^\dagger c_{k_yk_z \eta
  \tau {n}1} -c_{k_yk_z\eta \tau (n+1)\bar 1}^\dagger c_{k_yk_z\eta \tau n1} \nonumber\\
 & +  2 i \sin({k_ya_y})c_{k_yk_z\eta \tau n \bar 1}^\dagger c_{k_yk_z\eta \tau n 1}]+
  H.c. \nonumber, \\ 
  &H_{tk_yk_z} = t \sum_{n\sigma \left<\tau \eta;\tau'
  \eta'\right>}  e^ {i\phi_{\tau \eta \tau'
  \eta'}} c_{k_yk_z\eta' \tau' n\sigma}^\dagger c_{k_yk_z\eta \tau n \sigma}, 
  \label{tb}
\end{align}
where again the last sum runs over  neighboring layers and $\bar \alpha_\tau$
is the spin-flip hopping amplitude, related to the physical SOI parameter by
$\bar \alpha_\tau = \alpha_\tau /2a_y$ (assuming $a_x=a_y$) and to the SOI
energy by $\bar E_{so,\tau} = \bar
\alpha_\tau^2/t_0$~\cite{bena_majorana,Rainis2013,epb2}. Here,
$\phi_{\bar 1 1 1 \bar 1} = - \phi_{ 1 \bar 1 \bar 1 1} =k_z a_z$, otherwise,
$\phi_{\tau \eta \tau' \eta'}=0$.
The lattice constant in the $i$ direction  is $a_i$ with $i=x,y,z$.
The operator $c_{k_yk_z\eta \tau n \sigma}^\dagger$ is an annihilation operator
acting on electron with momentum $k_y$ ($k_z$) in the $y$ ($z$) direction and
with spin $\sigma$ located at the point $x=na_x$ along the $x$ direction of the
$\eta \tau$ layer. Our numerical results confirm the strong TI phase, see
Fig.~\ref{fig:num} and Supplemental Material (SM)~\cite{SM}. We again observe
the single anisotropic Dirac cone, where the accidental degeneracy at $k_z=0$
described before is lifted by a slight detuning of the chemical potential or
due to higher order tunneling terms not taken into account in the linearized
approximation~\cite{SM}.

\textit{Conclusions.} We introduced a coupled-layer approach to
construct a strong 3D TI,
where the building blocks are non-topological Rashba 2DEG layers. We showed that the bulk spectrum becomes
gapped, with the gap being proportional to the tunnel coupling $t$ between the
layers---a parameter that can be experimentally tuned. Additionally, any 2D boundary
hosts gapless helical surface states. We calculated the dispersion relation
of these surface states  and found a single Dirac cone at $\bm
k=0$, which together with the bulk gap constitutes a hallmark of a strong 3D
TI~\cite{HasanRMP}.

We acknowledge support from the Swiss NF and NCCR QSIT.

\bibliography{top}
\onecolumngrid

\author{Luka Trifunovic, Jelena Klinovaja, and Daniel Loss}
\affiliation{Department of Physics, University of Basel, Klingelbergstrasse 82,
CH-4056 Basel, Switzerland}

\bigskip 

\begin{center}
\large{\bf Supplemental Material to `From Coupled Rashba Electron and Hole Gas Layers to 3D Topological Insulators' \\}
\end{center}
\begin{center}
  Luka Trifunovic,$^{1,2}$ Jelena Klinovaja,$^1$ and Daniel Loss$^1$\\
$^1${\it Department of Physics, University of Basel, Klingelbergstrasse 82, CH-4056 Basel, Switzerland}\\
$^2${\it Dahlem Center for Complex Quantum Systems and Physics Department,\\Freie Universit\"at Berlin, Arnimallee 14, 14195 Berlin, Germany}
\end{center}

\twocolumngrid

\appendix
\section{Details of the analytical calculation}
\begin{figure}
  \includegraphics[width=\columnwidth]{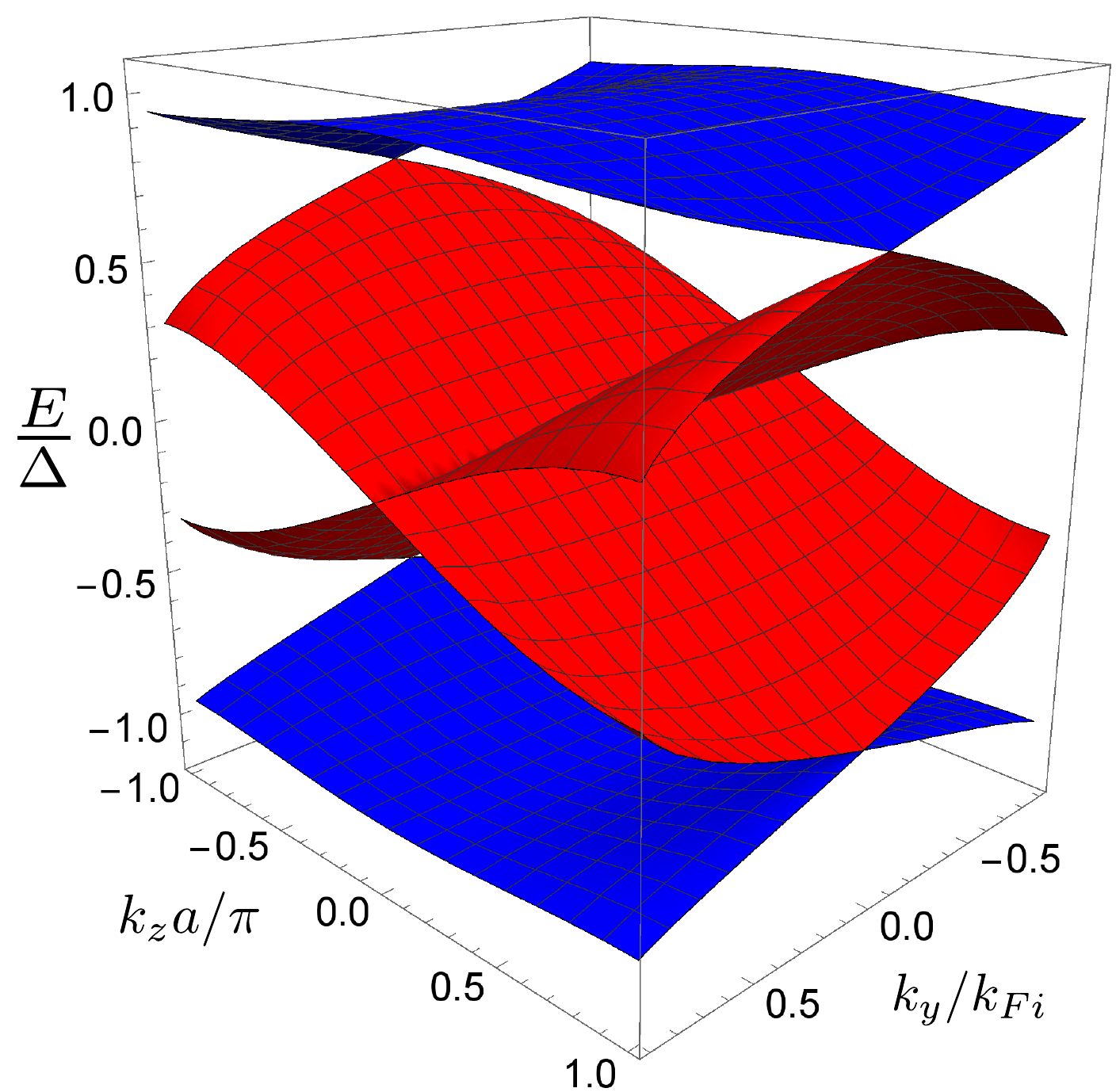}
  \caption{Dispersion relation of the surface states localized in the
    $yz$-plane for $k_{Fe}/k_{Fi}=3$, obtained analytically from
    Eq.~(\ref{eq:dispersion}) of the main text with $E/\Delta=\cos\Omega$. We
    plot $k_y$ up to value of $0.9k_{Fi}$, the concreate range of validity of
    the dispersion depends on the value of $t$ and is give below the
    Eq.~(\ref{eq:linearization}) of the main text.}
  \label{fig:yzdispersion}
\end{figure}

In order to obtain the spectrum of the surface states we fix the parameters
(including the energy inside the gap) and find the eight decaying eigenstates of
the Hamiltonian. Using Eq.~(\ref{eq:linearization}), we express them in the
basis of the original fermionic fields $\Psi(x)$, leading to eight eight-spinor
solutions $\Phi^j(x)$ with $j=1,\dots,8$, and construct a $8\times8$ Wronskian
matrix $W_{ij}(x)=[\Phi^j(x)]_i$. The equation $\det W(0)=0$ gives the spectrum
of the surface states~\cite{KSL2012}. We note that for $k_y\neq0$, the interior
and exterior branches have different velocities in $x$-direction. After
substituting $E=\Delta\cos\Omega$ with $\Omega\in[0,\pi]$ and assuming
$\theta_e,\theta_i\in[0,\pi/2)$ we obtain
\begin{align}
  \label{eq:Wronskian}
  \det
  W(0)=\frac{\upsilon_2e^{ik_za/2}}{\upsilon_1}&\left[4\sin^2(k_za/2)\cos^2\theta_e\cos^2\theta_i\right.\\
  &\left.-(\cos\theta_e+\cos\theta_i)^2\sin^2(2\Omega)
  \right].\nonumber
\end{align} 
Thus, the dispersion from Eq.~\ref{eq:dispersion} in the main text, shown in
Fig.~\ref{fig:yzdispersion}, is obtained.

\section{Detuning of the chemical potential}\label{sec:app_detuning}
In this section we show that any perturbation of the chemical potential in one
of the layers lifts the degeneracy for $k_z=0$ which is depicted in
Fig.~\ref{fig:yzdispersion}. In order to demonstrate this explicitly we assume a
detuning $\delta\mu$ of chemical potential in the first layer. For each value of
$k_y$ there is twofold degeneracy  which is lifted by such a perturbation. After
performing the perturbation expansion in lowest order within the twofold
degenerate subspace we arrive at the following dispersion relation

\begin{align}
  E(k_y,k_z=0)&=\delta\mu
  f\left(\frac{k_{Fe}}{k_{Fi}},\frac{t}{k_{Fi}\upsilon}\right)\frac{k_y}{k_{Fi}},
  \label{eq:dispersionP}
\end{align}
where we assumed $\upsilon_1=\upsilon_{\bar1}=\upsilon$ and $k_y\ll k_{Fi}$. The
function $f(x,y)$ is given by ($x>1$)

\begin{align}
  \label{eq:fxy}
  f(x,y)&=\frac{(x^2-1)^2
  }{8x( (x^2-1)^2+4y(x^2+1))}\sqrt{1 +\frac{4y^2}{(x-1)^2}}\nonumber\\
  &\times\sqrt{(
  2y(x-1)^2-4y^3)^2+(x+6y^2)^2}.
\end{align}

Since our analysis is valid for $t\ll k_{Fi}\upsilon$ [we took only resonant
terms into account in Eq.~(\ref{eq:Hnt})] we can further simplify the above
dispersion by expanding for small $t/(k_{Fi}\upsilon)$ and arrive at
Eq.~(\ref{eq:dispersionDirac}) of the main text.

\section{Numerical calculation of 2D surface states spectrum}
\begin{figure}
  \includegraphics[width=\columnwidth]{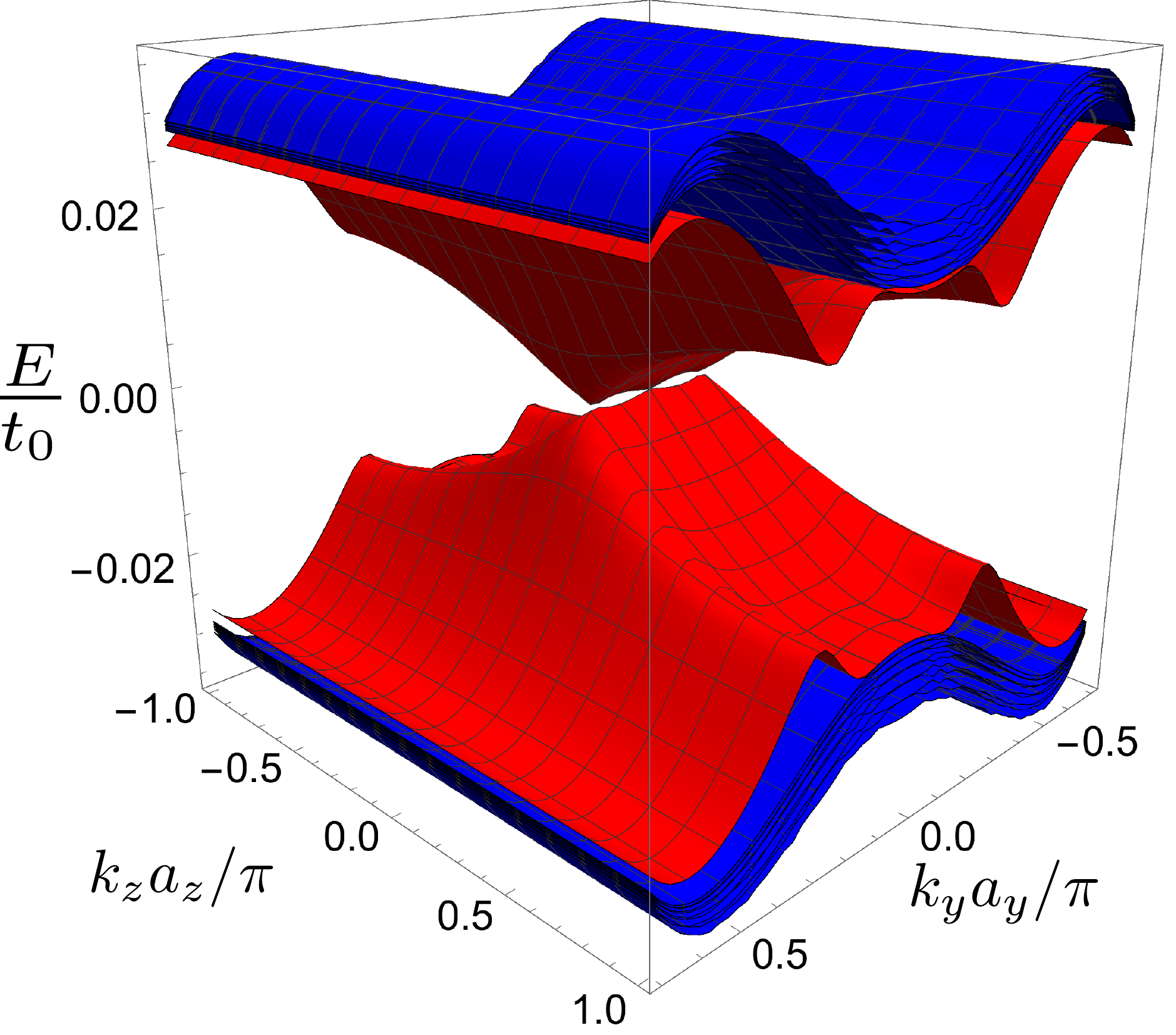}
  \caption{The dispersion relation of the surface states localized in the
    $yz$-plane obtained numerically in the tight-binding model. The parameters
    of the system take the following values: $\bar \alpha_{1}=0.3 t_0$,
    $\bar \alpha_{\bar1}=0.55 t_0$, $\vert\mu_1\vert=\vert\mu_{\bar1}\vert=0.2125 t_0$, and
  $t=0.05 t_0$.  }
  \label{fig:dispersion1}
\end{figure}
In this section we compare our numerical to analytical results and additionally
give more details about the numerical results. Our analytical results are valid
for $t\ll\vert\mu_\tau\vert,E_{so,\tau}$, and in this limit we obtain the
degeneracy for $k_z=0$, see Fig.~\ref{fig:yzdispersion}. This degeneracy is
lifted linearly in $k_y$ for $k_y\ll k_{Fi}$ as shown in the main text. Our
numerical tight-binding simulation confirms all these features, see
Fig.~\ref{fig:dispersion1}. Namely, around $k_y=0$ the degeneracy is linearly
lifted since for the tight-binding model it is very difficult to tune the sizes of
the Fermi surfaces to be the same across the layers. Additionally, we find that
there is a remaining degeneracy at $k_y\sim k_{Fi}$ ($k_{Fi}\sim0.2\pi/a_y$ for
the parameters in Fig.~\ref{fig:dispersion1}).

\begin{figure}
  \includegraphics[width=\columnwidth]{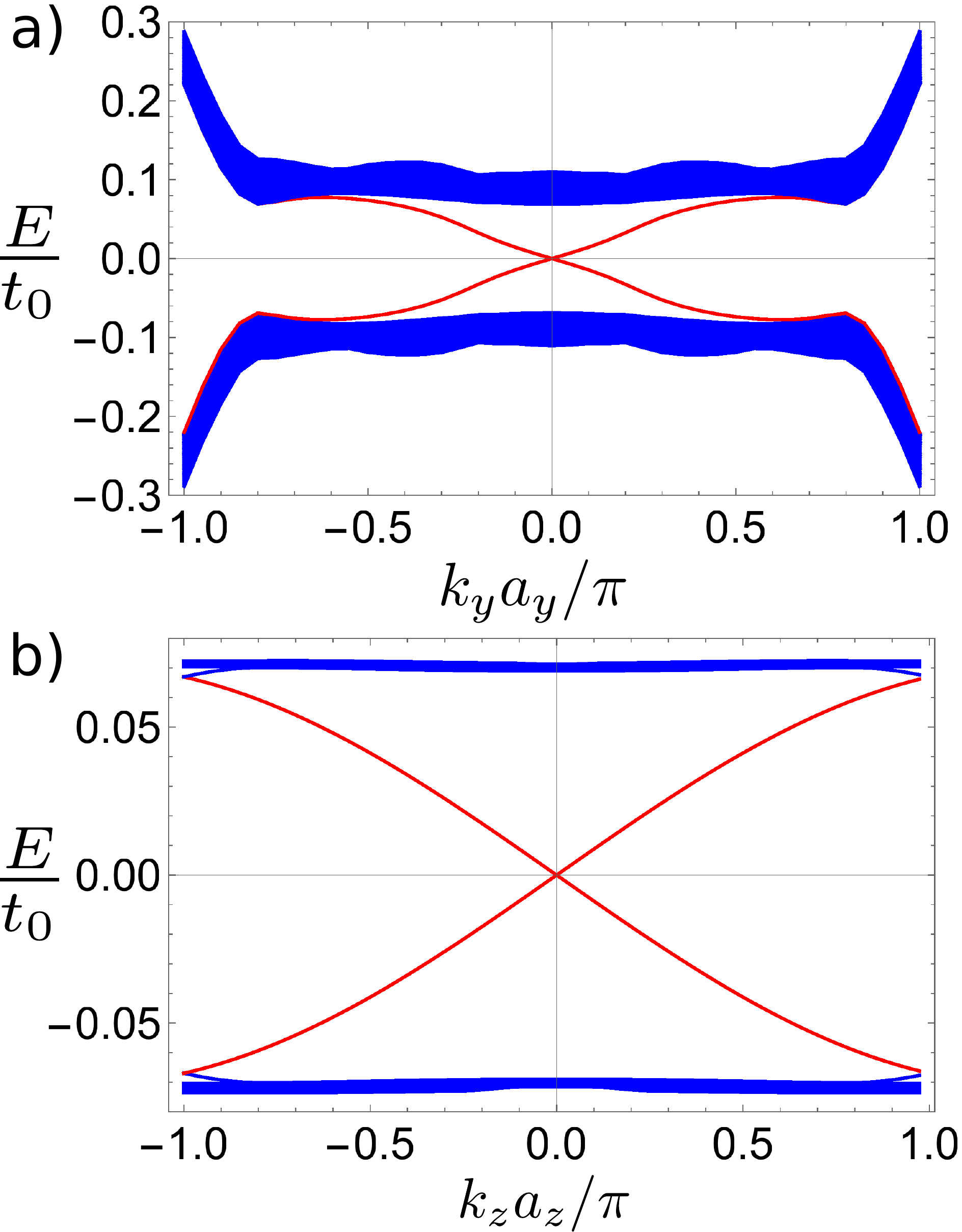}
  \caption{Panel a) [b)] show the $k_z=0$ [$k_y=0$] cut of the dispersion
    relation of the surface states localized in the $yz$-plane obtained
    numerically in the tight-binding model. The parameters of the system take
    the following values: $\bar\alpha_{1}=0.3 t_0$, $\bar\alpha_{\bar1}=0.55 t_0$,
    $\vert\mu_1\vert=\vert\mu_{\bar1}\vert=0.2125 t_0$, and $t=0.1 t_0$.}
  \label{fig:cuts}
\end{figure}

We found that increasing the tunnel coupling between the layers, above the
limit where the linearization works $t\sim E_{so,1}$, the  $k_y=0$ degeneracy
gets completely lifted and one obtains a single Dirac cone at $k_y=0$, see
Fig.~\ref{fig:num} of the main text. Additionally, in Fig.~\ref{fig:cuts}a
[Fig.~\ref{fig:cuts}b] we plot the cuts $k_z=0$ [$k_y=0$] of the dispersion
relation which show that there is no additional structure inside of the Dirac
cone. The Fig.~\ref{fig:cuts}a shows the behaviour of the surface states within
the whole Brillouin zone  from where it is seen that the dispersion relation curve of
the surface state does not bend down.

\end{document}